# Observation of the Ion-Ion Instability and its Suppression Mechanism in a Dusty Double Plasma Device

## Introduction

The spontaneous self-excited oscillation of random or organized motion in global or local approximation is a general property of dusty or complex or colloidal plasma system. This leads to some kinds of instabilities in presence of small size dust grains, which are similar to classical plasma instabilities without dust. Till now several dusty plasma instabilities have been observed experimentally [1 – 4]. These instabilities are of basic interest from the point of view of their collective processes of dynamical behavior. The main features of the dust instabilities are long characteristic times of their development, small changes of phase velocities and hypothetically variable charge fluctuations [5, 6].

The presence of charged dust grains can have a strong influence on the characteristics of the usual plasma wave modes with immobile dust background. When one considers frequencies well below the ion/electron plasma frequency ranges, new modes appear in the dispersion relation derived from either kinetic or fluid equations for three species consisting of ions, electrons and charge dust grains. A first observation of the self-excited dust acoustic wave instability of variable dust number density was carried out in a RF capacitive discharge low-pressure plasma device [7].

On the other hand, the DAW instability in the strata of DC glow discharge plasma was observed [8] by Milotkov *et al.*, which was later interpreted as dust ion streaming instability [9]. The existence of a low frequency dust wave mode in weakly coupled dusty plasma was theoretically predicted by [10] and then considered by many others. This mode is generally called dust acoustics wave because of its acoustic like dispersion relation in the limit of long wavelength. However, for strongly coupled dusty plasma, the behavior of the oscillation changes. There are few investigations [11, 12] of longitudinal and shearing type of waves in the limit of crystalline dusty plasma assumptions. They are usually termed dust lattice wave (DLW).

Instabilities in plasma can be considered as discrete phenomena due to internal feedback mechanisms, which are supposed to be governed by instantaneous coupling between boundaries [13]. On the other hand the suppression mechanism of various kinds of internal feedback instabilities generated in magnetized and un-magnetized plasma systems have been studied quite often [14 - 16]. Keen *et al.* [14] investigated the suppression of the ion acoustic instability by density perturbations in magnetized plasma under the condition when the perturbation frequency is close to and far away from the ion acoustic instability. They termed it synchronous and asynchronous suppression, respectively. In most of the cases the Van der Pol equation was mainly used for explaining the suppression mechanism of the instabilities. However, the use of the Van der Pol equation was not always successful in the explanation of the suppression of instabilities. With the assumption of an external feed back term in the Van der Pol equation, Nakamura (1985) tried to explain the suppression of the electron plasma instability [16].

In this paper, we present experimental observations for the evolution of the ion-ion instability and its different characteristics in a dusty double plasma device. Following this brief introduction, the experimental set up and the technique of measurements are described in section 2. Experimental results and discussions are briefed in section 3, while section 4 contains the conclusion of this paper.

**Experimental Setup**

Experiments were performed in a dusty double plasma device [17] equipped with multi dipole magnets for surface plasma confinements. The schematic of the device is shown in Fig. 1. The plasma inside the system is separated into two sections, named as source and target, by a stainless steel mesh grid of 81% optical transparency, which is kept electrically floating throughout the experiment. The dust dispersing chamber consists of an ultrasonic vibrator coupled to a dust reservoir and is mounted accordingly between the two magnetic bars at the top of the target chamber and very close to the separation grid.

The whole chamber is evacuated down to $(1.4 \sim 2.6) \times 10^{-6}$ mbar by a turbo molecular pump backed by a rotary pump. Argon (Ar) gas is introduced into the system under continuous pumping conditions at a working pressure of $4 \times 10^{-4}$ mbar. Plasmas in the source and target chamber are produced independently by DC discharge between the filaments and the magnetic cages. The discharge currents and voltages are set to 60 mA and 60 V, respectively, for the whole set of experiments. Plane Langmuir probe (LP) of 5 mm diameter and retarding potential analyzer (RPA) of 2.2 cm diameter are used to measure the plasma parameters and their fluctuations along the axis of the system. Fluctuations. These fluctuations are basically perturbed components of electron saturation current, and their frequency spectra are analyzed with the help of a spectrum analyzer. The RPA is used to measure the distribution functions of the ions, which are mainly effective for producing the ion-ion instability. Typical plasma parameters in the device are: Electron density $N_e = 10^7 \sim 10^8$ cm$^{-3}$, effective electron temperature $T_{eff} = 1 \sim 1.5$ eV and ion temperature $T_i$ 0.1. However, there are changes of the electron and ion saturation currents due to the introduction of dust particles inside the system, which are measured by the Langmuir probe and the RPA, respectively.

Dust particles in the form of glass beads of an average radius of 10 μm are used under these experimental conditions. To measure the dust density inside the system, a solid-state semiconductor laser of a wavelength of 630 nm and a power of 4 mW is used. Under this condition, the dust density ($N_d$) inside the system is measured by the relation,

$$I = I_0 \exp(-\pi r^2 L N_d), \qquad (1)$$

where, $I$ and $I_0$ are the measured laser intensities with and without dust, respectively, $r$ is the average radius of the dust grains and $L$ is the total length of the dust column. The measured dust density inside the system varies from $10^3$ to $10^5$ cm$^{-3}$. When the dust grains enter the plasma, they are charged by plasma electrons and equilibrium charge state is reached within very short timescale.

From the change of the electron and ion saturation currents measured by the Langmuir probe and the RPA, due to the introduction of dust particles, the experimental value of $\delta = \dfrac{N_{i,d}}{N_{e,d}}$ can be estimated and values of which is varied from 1.0 to 1.8 under our experimental conditions.

The amount of charge gained by the dust grains can be estimated by knowing the values of $\delta$ and $N_d$. Under these experimental conditions the dust charging time scale is very small (of the order of micro seconds). The charge on the dust grain is related to its surface potential ($V_s$) by $Q = CV_s$, where $C = 4\pi\varepsilon_0 V$ is the capacitance. The values of $Q$ obtained from both of these methods are within the experimental errors. Hence, the charge acquired by the dust grains under these experimental conditions are found to be $(-Q/e) \sim 10^5$, when $N_d < 10^3$ cm$^{-3}$, while this value is reduced to much smaller value for $N_d \geq 10^5$ cm$^{-3}$.

**Experimental Results and Discussions**

The ion-ion instability is excited in the system by injecting an ion beam into the source section of the device with a velocity $v_b \cong 1.9\ C_s$ (with $T_{eff} = 1.02$ eV and $T_i$ 0.1 eV). Here, $C_s$ is the ion acoustic velocity used for normalization. Frequency spectra of this kind of instability at various axial position of the target chamber, starting very close to the separating grid ($\sim z/\lambda_D = 40$), are shown in Fig. 2 (a). The amplitude of this instability is found to be maximum at $z/\lambda_D = 100$ from the separating grid, where $\lambda_D$ is the Debye length of the system without dust. It is important to note that the oscillation of this kind of instability is sustained only within $z/\lambda_D = 180$ from the grid. The oscillations of the higher harmonic, whose fundamental frequency and spectral width are almost 300 kHz and 150 kHz, respectively. Here, the width is considered as half maxima full width of the spectrum. Going away from the separating grid, the width of the spectrum broadens a lot without any change of the peak frequency for a fixed value of $v_b/C_s$. As $v_b/C_s$ increases or decreases further, the peak frequency changes accordingly, which indicates that the observed oscillations are standing waves, and the frequency change is due to the change in the mode number of the oscillations.

It is well known that when ion beams are injected into the plasma, two types of instabilities are excited. One is due to the inverse Landau damping of the ion beam and the other is due to the two-stream instability [18]. The growth rate of the instability due to the inverse Landau damping is very small (of the order of $10^{-3}$) compared to that of the two-stream instability. Hence, the instability excited in this case is basically the ion-ion instability. Various methods have been employed [18, 14, 19] to understand this kind of instability properly both theoretically and experimentally and also its suppression mechanism. The ion-ion instability is completely suppressed by RF pumps for $f_{ex}/f_{ii} = 1$ and $3 - 4$ [19], where $f_{ex}$ is the external frequency applied by the RF pump, and $f_{ii}$ is the ion plasma frequency, respectively. However, no one has introduced the characteristics of this kind of instability in a dusty plasma environment, where the dust density varies from $10^3$ to $10^5$ cm$^{-3}$. Frequency spectra of these instabilities in the presence of dust $N_d$ (or $\delta = 1.2$ and $1.6$) are shown in Fig. 2 (b, c) at different axial positions from the grid. The peak frequency of this instability decreases very slowly with the introduction of $N_d$ and the position of the maximum shifts very closely to $z/\lambda_D = 80$. It is also found that the spectral width increases considerably with the increase of $N_d$ and the peak frequency initially increases compared to the no-dust case, then decreases and then finally becomes zero at $N_d = 2.5 \times 10^5$ cm$^{-3}$, which is termed as critical dust density ($N_{dcr}$). The qualitative physical mechanism of the ion–ion instability in the presence of charged dust grains is described in detail in the later part of this report. The typical behavior of the ion current and its corresponding energy distributions are shown in Fig. 3 at different positions for $N_d = 0$ and $v_b/C_s = 1.9$, i.e., in the presence of the ion-ion instability measured by the directional retarding potential analyzer. Experiments have also been done in presence of different gas compositions like He, $H_2$, Xe and Ne. However, no remarkable changes are being observed.

Since the energy of the ion beam is very small ($\cong 1.7$ eV), it is very difficult to detect the beam energy by an RPA due to its resolution effect. However, there are some changes observed in the current and distribution function after numerical calculation of the first derivative of the respective *I-V* characteristics. The variations of the ion current and its corresponding energy distribution function at $N_d$ ($\cong 2.5 \times 10^5$ cm$^{-3}$ and $\delta = 1.8$) for different axial positions are shown in Fig. 4. In each case the energy distribution function is calculated numerically from the first derivative of the *I-V* characteristics. For $V_s = 1.75$ V, the beam energy $E_b = 1.67$ eV, the ion-ion instability appears. The disappearance of such a low energetic beam (Fig. 4) in the presence of a high critical dust density (i.e., $N_{dcr} = 2.5 \times 10^5$ cm$^{-3}$ or $\delta = 1.8$) is due to the scattering and extinction effect. It has been observed experimentally that for $v_b = 0.5\ C_s$, the wave velocity of the instability is smaller than the ion acoustic wave velocity in a normal two-component (Ar) plasma system [19]. Again, for $v_b/C_s$ between 0.5 and 2.0, the signal converts into the ion-ion instability. However, for $v_b/C_s$  2, the ion-ion instability disappears, and with a further increase of $v_b/C_s$ ($\cong 3$, 4 eV), the wave velocity of the fast mode, transformed from the ion acoustic mode, increases.

The ion distribution function obtained from the retarding potential analyzer gives a more reliable measurement of the beam velocity as long as the two peaks of the distribution function (i.e., background plasma ions and beam ions) are clearly distinct. This occurs when the beam energy is greater than the electron energy. At smaller velocities, the two peaks merge due to the plasma instability and also due to the deviation of low energetic ions because of its collision with charged dust grains near the grid sheath region. Again due to the introduction of dust particles into the system in the defined region, as explained in the experimental setup, the ion distribution function starts to change its behavior. This also corresponds to a small but negligible change of the ion-ion instability frequency. As the low energy ion beam becomes broader [20] in the presence of high dust density ($N_{dcr} = 2.5 \times 10^5$ cm$^{-3}$), the two peaks of the distribution function disappear completely. However, it is also clear from the distribution characteristics that the beam energy broadens at this condition and consequently the ion-ion instability disappears completely. A further increase of $N_d$ corresponds to a significant broadening of the beam energy. For high $N_d$ (or $\delta$), we introduce a high beam velocity in order to observe the two peaks of the RPA *I-V* characteristics, but beyond the condition of the ion-ion instability. From the broadening or scattering of the low beam energy or from the change of the ion current distribution of the RPA in the presence of dust particles, the effective cross section of the dust particles can be estimated [20].

The values of $T_e/T_b$ for various $N_d$ and plasma compositions in the whole set of experimental conditions are obtained as follows: since the beam ions have a drifting Maxwellian distribution, the beam temperature ($T_b$) of the beam ions in the moving frame of reference with the beam velocity $v_b$ is different from the laboratory frame of reference $T_{bl}$. It can be obtained by the following equation

$$T_b = T_{bl} + 4E_b - \sqrt{(4E_b^2 + 4E_b T_{bl})}, \tag{2}$$

where, $E_b = Mv^2/2$, $M$ is the mass of the ions, and $v_b = \sqrt{\dfrac{kT_b}{M}}$. For $E_b \gg T_{bl}$, $T_b$ is given by $T_{bl}^2/4E_b$. Under our experimental conditions, for $v_b = 1.7\ C_s$, i.e., $E_b \cong 1.45$ eV, Fig. 3 indicates $T_{bl} = 0.25$ eV, then the value of $T_b$ is estimated to be 0.011 eV. However, for lower values of $v_b$, its value changes significantly.

To understand the dispersion characteristics of the plasma in the presence of the ion-ion instability and charged dust grains, plane waves are excited by introducing a continuous small amplitude ($\cong$ 20 mV peak to peak) sinusoidal signal of variable frequency to the source plasma. The oscillation propagates perpendicular to the grid, which divides the source plasma form target plasma. The dispersion relations are measured experimentally by interferometer technique. Typical raw data taken from the interferometer output, when the probe moves away from the grid in the target side, is shown in Fig. 5 at the resonance frequency ($\cong$ 275 kHz, one third of the ion plasma frequency) for different $N_d$, when the probe is swept away from the separation grid. The growth and damping rates are almost equal when the applied wave frequency is very close to the excited instability frequency. The maximum amplitude of this interferometer pattern is found at $z = 100\ \lambda_D$ for $N_d = 0$ ($\delta = 1$) and for lower values of $N_d$. However, with the increase of $N_d$ ($\delta > 1.3$), the amplitude maxima start to shift slightly toward the grid ($z = 80\ \lambda_D$). The distribution function measured by the RPA at $\delta = 1.8$ shows a clear broadening of beam energy, which leads to the disappearance of the ion-ion instability in the system. We have tried to increase the beam density at these conditions, i.e., $\delta = 1.8$, but the ion-ion instability did not appear again. This indicates the suppression of this kind of instability at high values of $N_d$ (or $\delta$). The strong scattering or extinction of a low energetic beam in the presence of dust generally leads this kind of suppression mechanism.

The suppression mechanism of these kinds of instability has been discussed in a normal two-component plasma system [19]. There are two different kinds of suppressions namely synchronous and asynchronous suppression depending on the external frequency applied to the system. This kind of suppression mechanism is tentatively explained theoretically on the basis of the forced Van der Pol equation as [19],

$$\frac{d^2N}{dt^2} - (\alpha - 2\beta N - 3\gamma N^2)\frac{dN}{dt} + \omega_0^2 N = \omega_0^2 B\sin\omega t, \tag{3}$$

where, $N$ is the density perturbation, $\alpha$ is the linear growth, $\beta$ and $\gamma$ are nonlinear saturation coefficients, $\omega_0$ is the characteristic angular frequency, $\omega$ is the angular frequency of the externally excited wave and $B$ is the amplitude of the external signal. The above equation has been used for synchronous suppression, and for the asynchronous suppression $\omega_0$ has been replaced by $\omega$. If the equation of continuity, equation of motion and Poisson equation are considered as basic equations then the right hand side of the forced Van der Pol equation has the above form. In their theoretical model a spatially uniform phenomenon was considered. But in the experiment this is not really true. In our experiment, the amplitude of the instability wave increases spatially up to $z/\lambda_d = 80$ to 100, depending upon the value of $\delta$ and then finally decreases away from the separation grid. In this condition, the broadening of the energy distribution of the beam ions [Fig. 4] due to the presence of dust particles, which are basically causing the suppression of the ion-ion instability, requires a modification of the Van der Pol equation. The coupling of the ion-ion instability with the externally excited wave transforms energy in order to suppress the instability, which is considered to be main physical mechanism behind the suppression.

The physics laid forth above for the suppression mechanism of the ion-ion instability offer qualitative arguments under our special experimental conditions. However, in order to have a clearer picture of this kind of instability in the presence of charged dust grains, we have to consider the trapping of beam ions by the highly charged dust grains.

**Conclusion**

This work deals mainly with the decay and growth rate of the ion-ion instability with charged dust grains. It is found that at a critical dust density the ion-ion instability is suppressed completely and the energy of the ion beam broadens. Many authors have tried to understand the suppression mechanism in two-component plasma with the help of the Van der Pol equation with a forced term on the right hand side of the equation. Both the synchronous and asynchronous suppression with the help of an external frequency applied to the system were considered in their papers. However, in our case the suppression is mainly due to the charged dust grains inside the system. Hence, a proper physical validation of the suppression of such a kind of instability in a charged dust environment is open for future considerations.

**Acknowledgement:**

One of the authors A. Sarma thanks Department of Science and Technology, Government of India for providing financial support under BOYSCAST Fellowship scheme.

**Figure captions:**

Figure 1: Schematic diagram of the experimental setup. L: Langmuir probe; E: Retarding potential analyzer; S: Source chamber; T: Target chamber; G: Separation grid; I: Ionization gauge; F: Filament; D: Dust particle reservoir; C: Dust particle collector; V: Ultrasonic vibrator; $V_s$: Power supply for source anode biasing voltage.

Figure 2 (a): Frequency spectra of the ion-ion instability measured with the help of a spectrum analyzer without dust detected by Langmuir probe. In this case $v_b/C_s$ = 1.9, $\delta$ = 1.0 and data taken for the distance $z$ = 40 $\lambda_D$ from the separation grid.

Figure 2: Frequency spectra of these instabilities at different axial positions ($z/\lambda_D$) from the grid.
(b) $\delta$ = 1.2 (c) $\delta$ = 1.6

Figure 3: Collector current profile measured by directional retarding potential analyzer at different axial positions *viz.* $z/\lambda_D$ = 40, 100 and 200 and $\delta$ = 1.0. Dotted curves represent $dI_c/dV$ of the respective curves.

Figure 4: Collector current profile measured by directional retarding potential analyzer at different axial positions *viz.* $z/\lambda_D$ = 40, 100 and 200 and $\delta$ = 1.8. Dotted curves represent $dI_c/dV$ of the respective curves.

Figure 5: Interferometer raw data taken by Langmuir probe swept away from the grid through a distance $z/\lambda_D$ = 40 to 200 for different values of $\delta$ at a frequency 275 kHz, which is close to peak of the excited instability frequency.

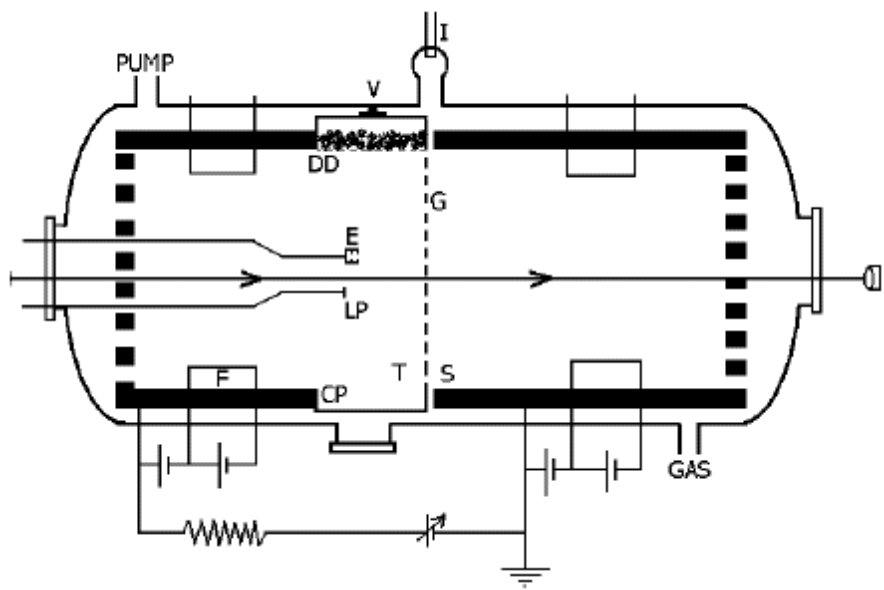

**Figure: 1**

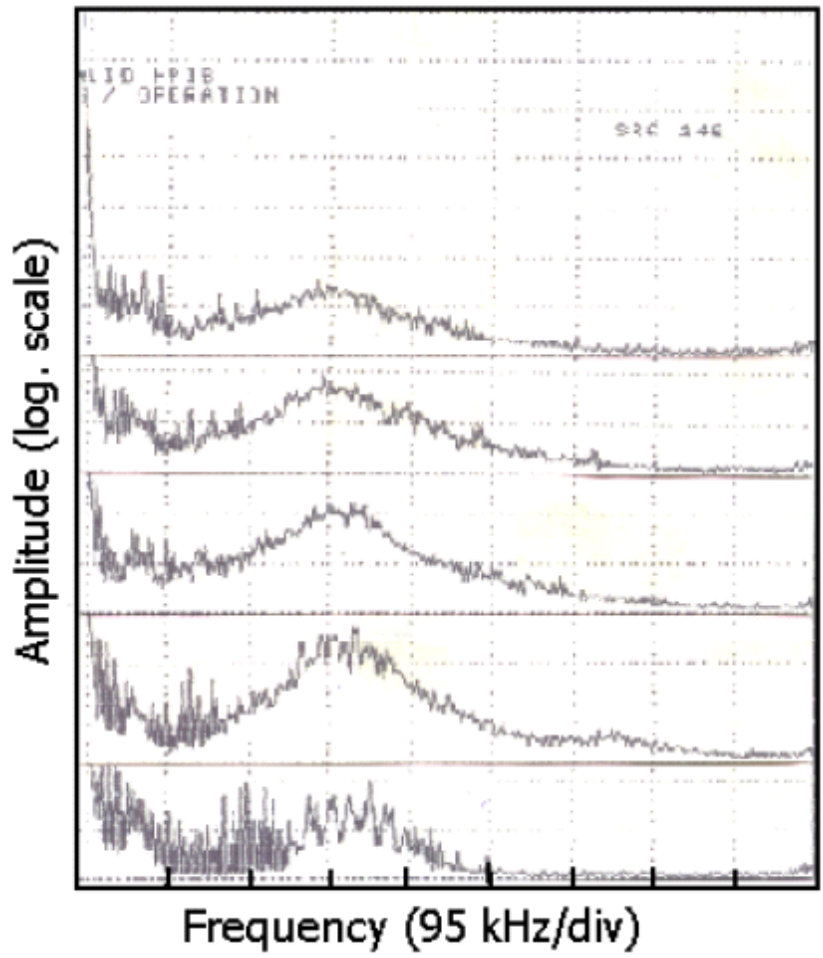

Figure: 2(a)

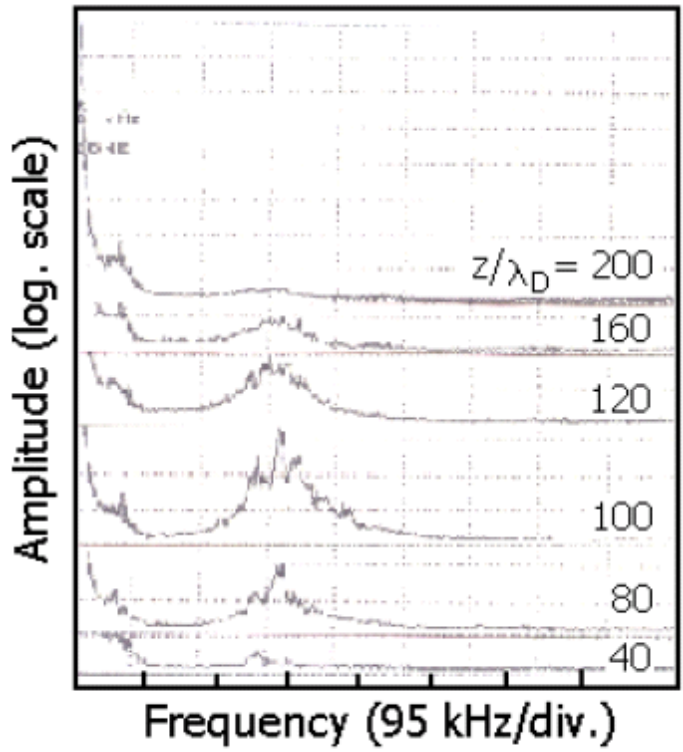

**Figure 2(b)**

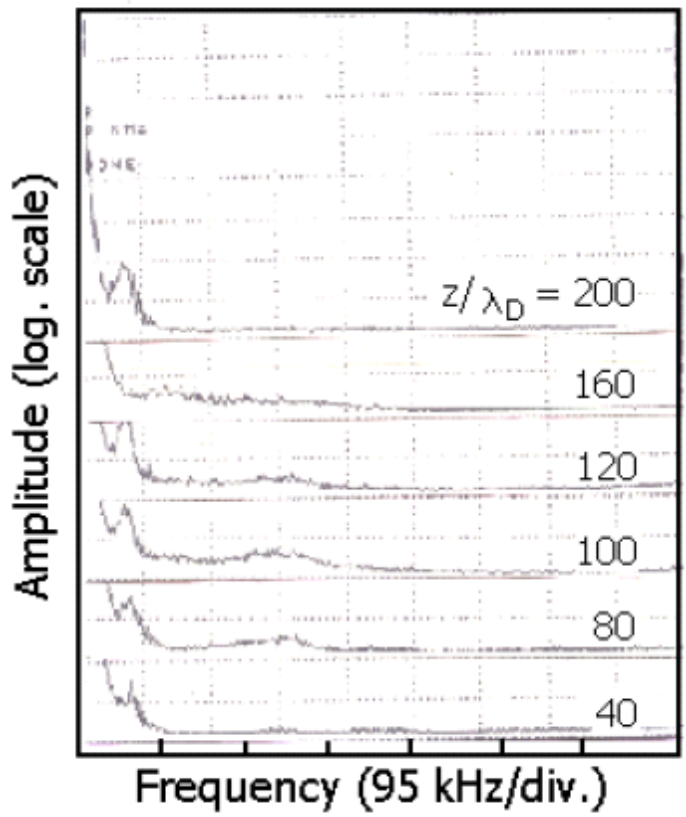

Figure: 2(c)

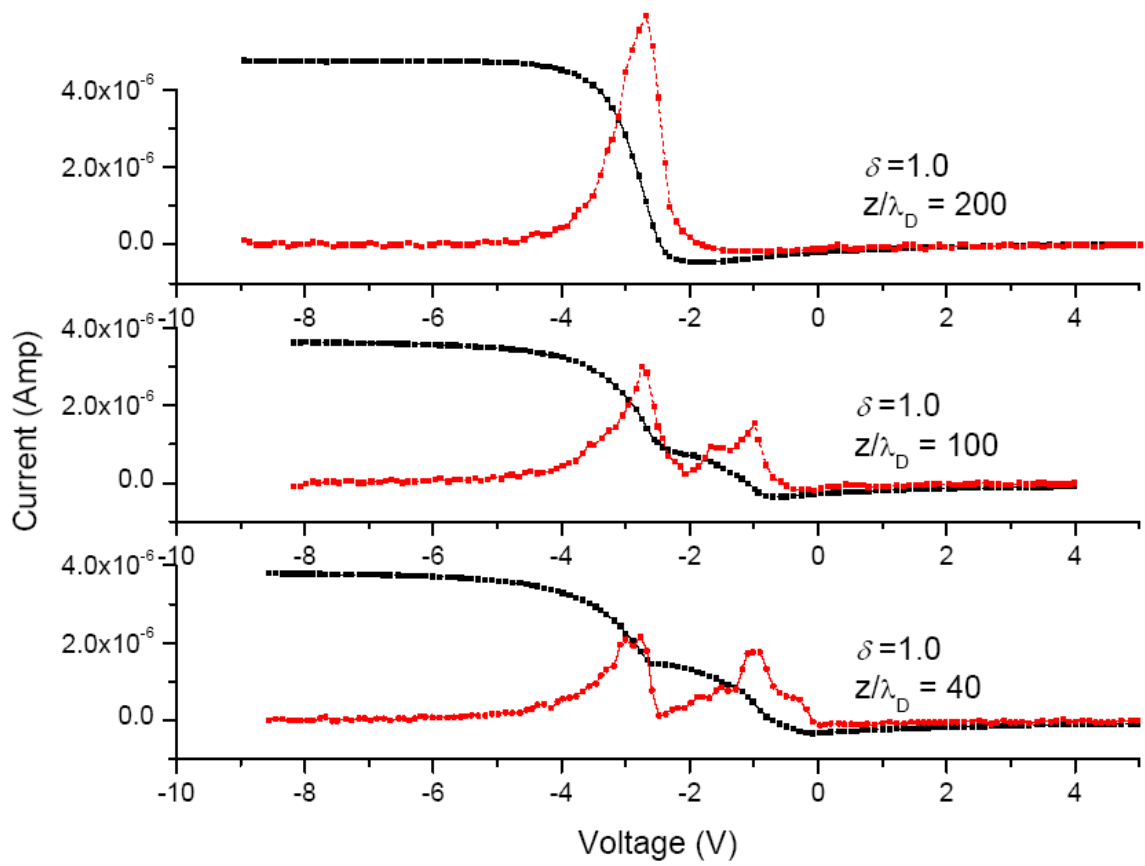

**Figure: 3**

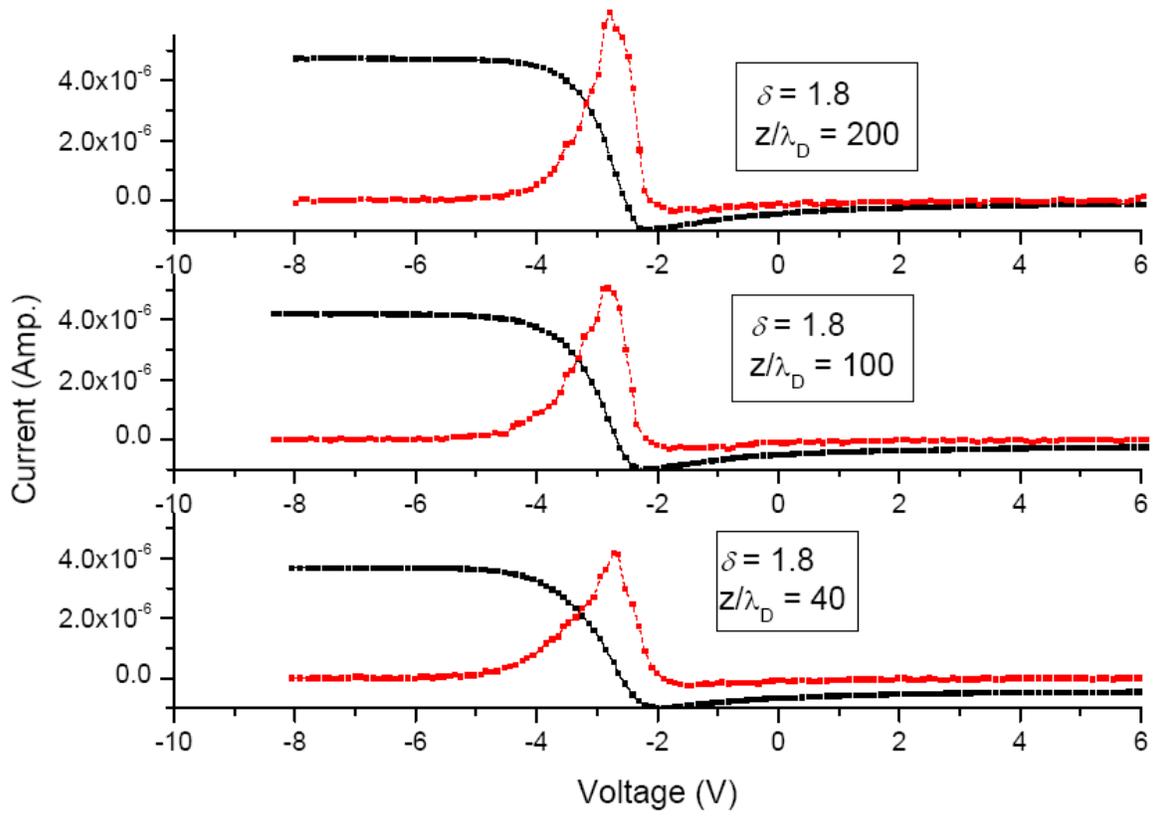

**Figure 4**

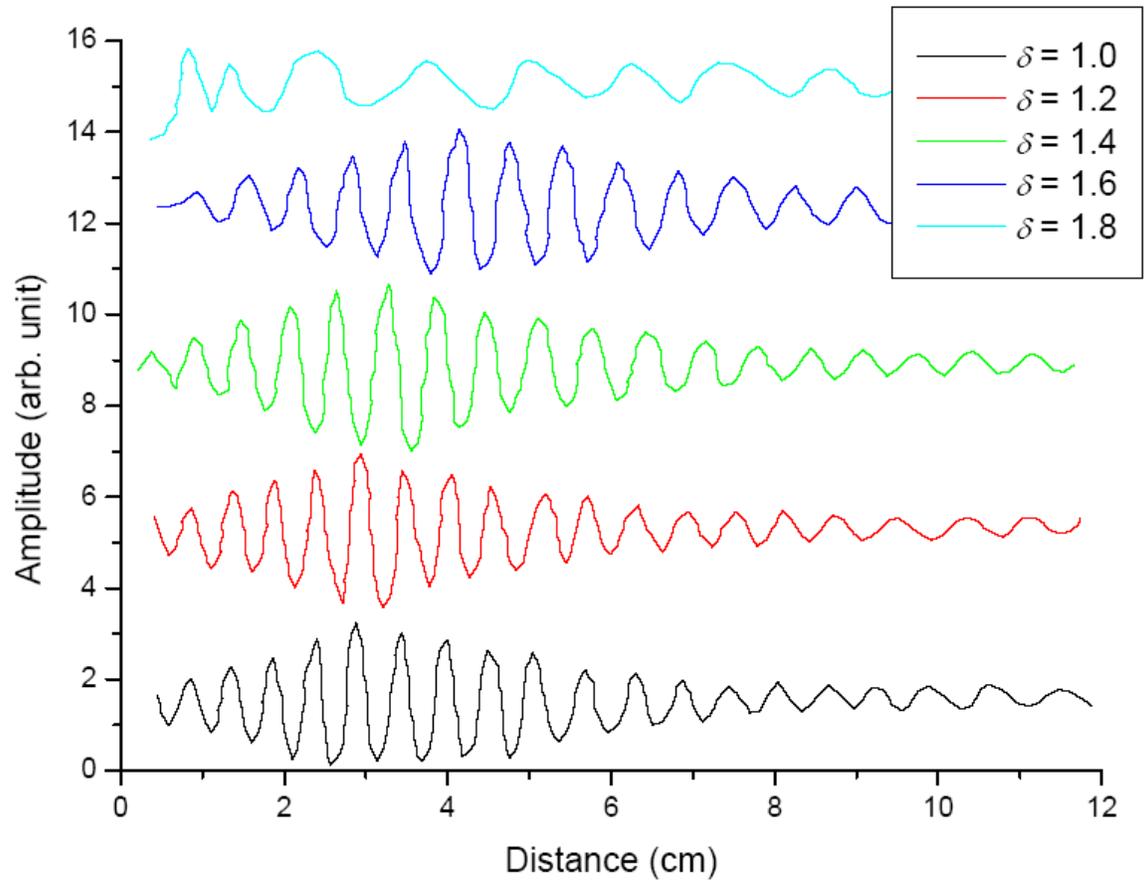

**Figure 5**